\documentclass[preprint2]{aastex}

\tighten

\shortauthors{Boroson}
\shorttitle{New Orientation Indicator}

\begin{document}

\title{A New Orientation Indicator for Radio-Quiet Quasars}

\author{Todd A. Boroson}
\affil{National Optical Astronomy Observatory\footnote{The National Optical
Astronomy Observatory is operated by AURA, Inc., under cooperative agreement
with the National Science Foundation.},       
P.O. Box 26732, Tucson, AZ 85726}

\vfill

\begin{abstract}

The velocities of the [O III] $\lambda$5007 and optical Fe II
emission lines, measured relative to the systemic redshifts of 2265 QSOs by
\citet{hu08}, show the signature of a disklike BLR structure with polar outflows.  
Objects with large 
[O III] outflows show no Fe II offset velocity and are seen pole-on.  
Objects with large Fe II inflow show no [O III] offset velocity and are seen 
edge-on.  This interpretation is supported by the morphology of the radio-loud
objects within the sample and by previous determinations of the geometry of
the broad and narrow line regions.  Analysis of the objects with neither
Fe II or [O III] velocity offsets, however, show that the two groups also differ in 
Eddington ratio, and, within this subset, corresponding groups with high and 
low Eddington ratio but with the opposite orientation can be identified.

Using these four subsets of the sample, the effects of orientation and Eddington ratio
can be separated, and, in some cases, quantified.  The changes in apparent continuum
luminosity and broad H$\beta$ width and strength suggest a model in which both continuum 
and H$\beta$ are emitted from the surface of the disk, which is less flattened in high
Eddington ratio objects.  The effects of orientation on the derived properties, black hole mass
and Eddington ratio, are significant, though not large.  The [O III] outflow appears to influence
the width of that line, as well as its centroid.

\end{abstract}


\section{Introduction}
It has long been recognized that AGN are not spherical.  Angular momentum
of the fueling material implies the development of a disk.  Radio jets are obviously bipolar. 
The obscuring material that drives the Type 1-Type 2 dichotomy is thought to be toroidal.  
In nearby objects, the narrow line region (NLR) is often seen to be biconical.  

However, the compelling case for a quantitative orientation indicator has been problematic.  
In radio-loud objects, it is believed that the degree of radio core dominance is indicative of
the orientation, in that bright cores represent the doppler-enhanced emission from viewing close
to the axis of the jet \citep{blandford79,wills86}.  Radio-loud objects represent a small fraction of AGN, though, and even 
within this fraction, the intrinsic radio properties vary with radio luminosity.  Thus, the 
identification of large samples of objects that are identical other than orientation is not straightforward.

For radio-quiet objects, the recent availability of very large and relatively complete samples 
of AGN from the Sloan Digital Sky Survey (SDSS) \citep{york00} make a statistical approach possible for the first time.  
While studies of samples of tens or hundreds of objects \citep{bg92,marziani03} have 
discerned patterns and correlations in the observed properties of QSOs, an understanding of
how to identify and separate the effects of the overall drivers -- both physical and geometrical -- requires
much larger samples.  Important steps in this direction have been taken by studies such as
\citet{richards02}, \citet{hu08}, \citet{ris11}, and \citet{decarli11}, all of whom find evidence 
for orientation effects among the properties that they investigated.  None of these studies, however, 
have produced a general technique for determining the orientation of a given object, or for identifying subsets of analogous objects differing only by orientation.

Despite the lack of an unambiguous orientation indicator, a picture of the geometric structure 
of active nuclei
has emerged \citep{urry95,gaskell09,crenshaw10}.  The region that emits the lower ionization broad lines is thought to be somewhat 
disklike.  Material in this disk may be flowing inwards, but there may also be rotation and 
turbulence.  The optical continuum may also be emitted from a disk, interior to the broad line
region (BLR).  These disks are 
assumed to be coplanar with the dusty torus that obscures our view of the continuum and BLR 
in Type 2 objects.  The higher ionization narrow lines are believed to arise in a biconical region
that is perpendicular to the disk and flowing outward.  Some fraction of the higher ionization 
broad lines may also participate in this polar outflow.  The lower ionization narrow lines 
probably arise over a very large region dominated by the potential of the host galaxy rather than
the central black hole.

In section 2, a diagram is constructed showing the relationship between the motions of the material that emits broad Fe II and that which emits narrow [O III].  This diagram is most obviously interpreted as indicating 
the orientation of two groups of quasars, one pole-on and one edge-on, within the general picture
described above.  Exploration of intermediate objects show that another factor, in addition to orientation,
is at work; this is identified as L$_{bol}$/L$_{Edd}$ or Eddington ratio (ER).  These intermediate objects can
be separated into subsets that match the pole-on and edge-on subsets in ER, but differ in orientation.  Section 3 presents the characteristic properties of each of these four subsets, allowing a separation of
the effects of the two factors.  Finally, section 4 discusses the implications of the orientation-driven
effects for determinations of black hole mass (M$_{BH}$) and ER.  Section 4 concludes with some
questions that could be answered by further development of this technique.

\section{The $V_{[O III]}$ vs. $V_{Fe II}$ Diagram}

The study by \citet{hu08} was aimed at understanding the properties of Fe II emission in quasars.  
They decomposed the spectra of 4037 z$<$0.8 quasars into continuum and 
multiple line contributions.  These authors measure luminosity, width, and velocity shift parameters
for H$\beta$ (broad and narrow separately), [O III] $\lambda$5007, Mg II $\lambda$2800, 
and [O II] $\lambda$3727.  Because of the range of redshift covered, the Mg II and [O II] lines 
are only measured in a subset of the full sample.
They also measure these parameters for the ensemble of Fe II
lines in the range $\lambda\lambda$4434-4684.  They devote considerable effort to testing 
and demonstrating the accuracy of their technique - in particular, the velocity shifts.

Their sample is drawn from the SDSS DR5 \citep{adelman07} quasar catalog 
\citep{schneider07} with the 
following additional restrictions.  They attempt to fit only objects having signal-to-noise
ratio (S/N) $>$ 10 
in the range $\lambda\lambda$4430-5550.  They then discard objects with poor fits to 
their model or with equivalent width of the Fe II emission (EW$_{Fe II}$) $<$ 25\AA\ . Finally,
they discard objects with large errors in either the width of the broad H$\beta$ or the velocity
shift of [O III] $\lambda$5007.  These restrictions, particularly the minimum EW$_{Fe II}$, should
be kept in mind for statistical conclusions drawn from their sample.

\citet{hu08} adopt the [O III] $\lambda$5007 line to define the systemic redshift because they
can measure its position in all objects in their sample.  However, $\lambda$5007 is known to
be blueshifted or to have excess emission on its blue wing in many objects \citep{zamanov02,boroson05}.  Shifts as large as several hundred km s$^{-1}$ are occasionally
seen.  For this study, all the velocities have been renormalized to the [O II] $\lambda$3727 
line, which is measured in 2265 of the objects.  Objects without a measurement of velocity shift for [O II] were discarded.

Figure 1 shows the [O III] velocity shift plotted against the Fe II velocity shift for the 2265 objects.  The random errors on the Fe II velocity shifts are about 100 km s$^{-1}$ for small shifts and rise to about 300 km s$^{-1}$ for the shifts of 3000 km s$^{-1}$.  The errors on the [O III] velocity shifts are about 25 km s$^{-1}$ for small shifts and rise to about 50 km s$^{-1}$ for the largest blueshifts, 700 km s$^{-1}$.  While the density of objects is highest around the origin, two tails are clearly seen.  About 37\% of the objects show almost no [O III] velocity shift, but significant Fe II redshift.  About 10\% of the objects show almost no Fe II velocity
shift, but significant [O III] blueshift.

\subsection{Effects of Orientation}

\citet{hu08} argue that the Fe II redshift is caused by infall of material in the outer part of the 
BLR.  They find that the Fe II lines are uniformly narrower than H$\beta$ and that the objects with
large Fe II redshifts show excess emission on the red side of the H$\beta$ line.  They also find 
that the Fe II redshifts are inversely correlated with ER.  They speculate that the Fe II
infall is driven by gravity and opposed by radiation pressure.  The rise in radiation pressure with
increasing ER decreases the inflow.

There is direct evidence supporting the idea that the blueshift of [O III] is indicative of a polar outflow.  Studies of very nearby objects \citep{ruiz05, crenshaw10} are able to resolve
the NLR sufficiently to map the spatial and kinematic distribution of emitting material and match
it to models of outflowing material and a biconical illumination pattern.  

In light of these explanations of the velocities of the Fe II and [O III] emitting material, the obvious 
conclusion is that the tails in the v$_{[O III]}$ vs. v$_{Fe II}$ diagram are indicative of orthogonal motions - Fe II in the disk and [O III] in the perpendicular polar direction.  In this case, the objects
with large Fe II inflow velocities are seen with the disk edge-on to our line of sight, and the
objects with large [O III] outflow velocities are seen with the disk pole-on to our line of sight.
Therefore, we define three regions, shown in figure 1 as pole-on and edge-on subsets, and a third
subset that includes objects with neither large Fe II velocities nor large [O III] velocities.  
The boundaries of these regions are arbitrary; they merely serve to isolate the two tails of the
distribution.  These boundaries result in 829 edge-on objects, 226 pole-on objects, and 1081 
intermediate objects.  Note that the term edge-on is used in a relative sense; an obscuring torus in the same plane as the Fe II emitting disk would limit actual viewing angles to be outside of a true edge-on value.

Some confirmation that this explanation is correct comes from the radio-loud objects in the 
sample.  Within the edge-on subset, there are 778 objects within the footprint of the FIRST radio
survey \citep{becker95}.  A search of the FIRST archive at these positions yielded 59 objects with R $>$ 10, where R 
is the ratio of 6 cm to 2500 \AA\ flux.  The FIRST images of each of these matches was inspected visually and classified as core, extended or resolved single source, core plus lobe, core plus 2 lobes, or 2 lobes with no core.  Thirty of these edge-on objects are not compact radio sources,
and of these, 17 show two lobes.  Similarly, the pole-on subset was analyzed.  Within that smaller
subset, there are 7 radio-loud objects, of which 5 are compact, one is a resolved single source,
and one has a core plus one lobe morphology.  Assuming that the radio source axis is 
aligned with the [O III] outflow direction, this is as expected; double lobed objects are only seen 
within the edge-on subset. The intermediate subset shows intermediate radio properties; of the 40
objects in this subset that are radio-loud, five show two lobes.  However, note that the radio-loud fraction of all these subsets is small, and the sample-wide characteristics apply to the radio-quiet objects as well.

\subsection{Effects of Eddington Ratio}

The argument that the tails in figure 1 isolate pole-on and edge-on subsets does not 
preclude the possibility that other factors play a role.  One that should certainly be considered 
is L$_{bol}$/L$_{Edd}$, or 
Eddington ratio.  Many trends in AGN characteristics, and, in particular, both of the effects 
that contribute to the v$_{[O III]}$ vs. v$_{Fe II}$ diagram 
have been related to Eddington ratio. \citet{hu08} argued
that as Eddington ratio increases, radiation pressure opposes and eventually prevents the
Fe II inflow. Similarly, \citet{boroson05} and \citet{zamanov02} both find correlations between the
objects with largest [O III] blueshifts and Eddington ratio.  More generally, Eddington ratio has been
proposed as a driver for the EV1 correlations \citep{bg92}, which include FWHM of H$\beta$ and 
EW of both Fe II and [O III].  

We begin with the hypothesis that the sequence of objects in the v$_{[O III]}$ vs. v$_{Fe II}$ diagram
is due only to orientation.  As one's viewing angle changes from edge-on to pole-on and passing 
through an intermediate range, the Fe II inflow velocity becomes apparently smaller, while the [O III]
outflow velocity becomes apparently larger.  Within the intermediate range, objects show neither 
motion strongly and cluster around the origin.  If orientation were the only factor, then properties of the objects
near the origin should be intermediate between the two tails.  Furthermore, relationships among those
properties should behave in a way that tracks the relationships between the two tails. 

The orientation-only hypothesis is disproven by figure 2, which shows the luminosity at 
$\lambda$5100 plotted against the
FWHM of H$\beta$ for the objects that have no significant offset velocity in either Fe II or [O III].  While the edge-on objects have lower 
luminosity and larger
line width than the pole-on objects, the putative intermediate objects show the opposite trend -- higher
luminosity objects have larger line width.  The correlation coefficient for these 1081 objects closest to the
origin is 0.35, indicating a probability of less than 10$^{-7}$ of no correlation.  Thus, our conclusion
is that this intermediate subset represents a mix of the low Eddington ratio objects that are pole-on
and the high Eddington ratio objects that are edge-on.

\subsection{Combining the Two Drivers - Orientation and Eddington Ratio}

The next step is to divide the intermediate subset into those objects that represent the pole-on counterparts of the edge-on, low Eddington ratio objects and those that represent the edge-on 
counterparts of the pole-on, high Eddington ratio objects.  To do this, note that the solid angle over which
objects appear pole-on is smaller than that over which they appear edge-on.  Counting objects in
the two tails of figure 1 and assuming that these tails represent objects that are within 30 degrees
of the preferred orientation, we calculate that approximately half of the intermediate objects 
should be the edge-on counterparts of the objects that show large [O III] outflows and half 
should be the pole-on counterparts of the objects that show large Fe II inflows.  

Because our interpretation of figure 2 is that the correlation seen between L$_{5100}$ and 
FWHM H$\beta$ is due to the mix of objects with two different sets of physical parameters, we
divide the distribution of objects in figure 2 with a line perpendicular to the best fit line (the bisector
of the two dashed lines shown), at a 
point that divides the distribution into two approximately equal subsets.  We choose to label the upper right end of the distribution as the low-Eddington pole-on objects because if the two parts of the intermediate objects were assigned in the
opposite sense, they would appear to get brighter when they were viewed edge-on than when
they were viewed pole-on.  However, both the division and the assignment are speculative and
based somewhat on a preconception about the structure and kinematics of the objects in the
sample.

\section{How Some Properties Vary with Orientation and Eddington Ratio}

Table 1 gives median properties for the four subsets.  All the properties are calculated from data in Table 2 in \citet{hu08}.  Note that the black hole masses are calculated using the values of $\sigma_{H\beta}$
rather than FWHM H$\beta$, and that the Eddington ratios are calculated using 9$\lambda$L$_{5100}$ 
as the bolometric luminosity.   Uncertainties in these characteristic 
values for the subsets are in the range of only a few percent, though it should be noted that the range of any
property within a subset is large, as can be seen in figure 2.  While some
quantitative conclusions are drawn below about the behavior of properties within and between subsets, we caution that (a) the subsets are drawn from a parent sample that has been chosen in a way that undoubtedly skews some
of these statistics and (b) our definitions of the subsets and how they are associated is arbitrary, though 
we believe that it all creates a consistent picture. Note particularly that the entire sample is assumed 
to be the sum of two separable populations, and this is unlikely to be the case.  In reality, it 
is doubtful that a sharp transition exists between objects that show an Fe II inflow if they are seen edge-on
and those that show an [O III] outflow if they are seen pole-on.  Also, our classification of objects as either edge-on or pole-on certainly weakens the apparent effects.

The behavior of luminosity and broad line width is particularly important as these observables are used to derive black hole mass and Eddington ratio.  Both of these properties behave qualitatively as expected 
given the results of previous studies of the general structure of the BLR and continuum emitting region.  Table 2 shows that L$_{5100}$ increases as objects are seen more pole-on: by a factor of 2 for high Eddington ratio objects and by a factor of 3 for low Eddington ratio objects.  H$\beta$ width is not strongly
dependent on orientation for high Eddington ratio objects, but for low Eddington ratio objects, it is larger
by about 15\% in those objects seen  edge-on. 

Consistent with the \citet{ris11} study, EW$_{[O III]}$ is smaller in pole-on objects than in edge-on objects, though the effect is much less than would be expected if [O III] emission were isotropic, and the decrease were due only to the increase in the apparent continuum luminosity.  Similarly, the [O III] 
outflows also seem to broaden the [O III] lines, greatly in the high Eddington ratio objects (by definition), but slightly even in the low Eddington ratio objects, while no such effect is seen in the [O II] lines.

\section{Discussion}

Overall, a picture consistent with these findings is as follows.  The continuum is emitted from the surface
of a disk and the BLR is somewhat flattened also.  At high ER, both of these regions are less flattened than at low ER.  The Fe II inflow, seen in low ER objects, is roughly in the plane of this disk, though it may be that in edge-on objects, we see it at a steep angle, but not zero degrees because of the coplanar obscuring torus.  Perpendicular to this disk is the biconical outflow of the narrow forbidden lines.  In high ER objects, we see the [O III] coming towards us when our view is pole-on.  Lower ionization species probably arise further out where the flow has decelerated.

Estimates of the effect of orientation on L$_{5100}$ and FWHM H$\beta$ allow an assessment of its impact on derived quantities M$_{BH}$ and L$_{bol}$/L$_{Edd}$.  Most of the effect appears to be on luminosity, where a factor of 2-3 between edge-on and pole-on objects translates into a shift in both M$_{BH}$ and
ER of 40\% 
to a factor of about 2.  The effect on line width, apparently significant only for low Eddington ratio objects,
translates into an increase of about 30\% in calculated M$_{BH}$ for edge-on objects, but in the opposite
sense to the luminosity effect.  It is possible that the line width effect is substantially larger than is seen in this sample, since the objects with very broad H$\beta$ lines are known to have relatively weak Fe II emission, and so have been preferentially excluded from the \citet{hu08} sample.   If the quantitative differences between subsets are used to correct each subset to an average orientation, the corrections
change the M$_{BH}$ and L$_{bol}$/L$_{Edd}$ numbers by a few tens of percent.  They bring the M$_{BH}$ values closer for similar objects at different orientations.  The high and low Eddington ratio subsets 
retain the same relationship; the high Eddington ratio objects have 2 - 3 times the Eddington ratio value
of the low Eddington ratio objects.

This work is a first step towards understanding orientation effects based on the inflow and outflow velocities in large samples.  Several questions that represent possible next steps are:

1. For objects without strong Fe II, can the asymmetry of the H$\beta$ line be used to identify objects that are seen close to edge-on?

2. Can a sufficiently complete and well-defined sample of objects be analyzed in this way to separate orientation-driven and Eddington ratio-driven trends with higher precision?

3. Does a correction to M$_{BH}$ based on orientation have the effect of reducing the scatter in the M$_{BH}$--$\sigma_{\star}$ relation or of clarifying the location of Narrow -Line Seyfert 1s in that diagram?

4. Can trends in other properties, for example, the shape of the H$\beta$ line, be shown to correlate with
orientation, in order to provide additional information about the structure or kinematics of the inner regions
of QSOs?

\acknowledgments

I thank Luigi Foschini and the organizers of the conference {\it Narrow Line Seyfert 
1 Galaxies and Their Place in the Universe}, held in Milan, Italy in April 2011 for the
invitation that led to this work.  I thank Hermine Landt, Guido Risaliti, and Alessandro Marconi for helpful discussions.

\clearpage

\begin{figure}
\epsscale{1.}
\plotone{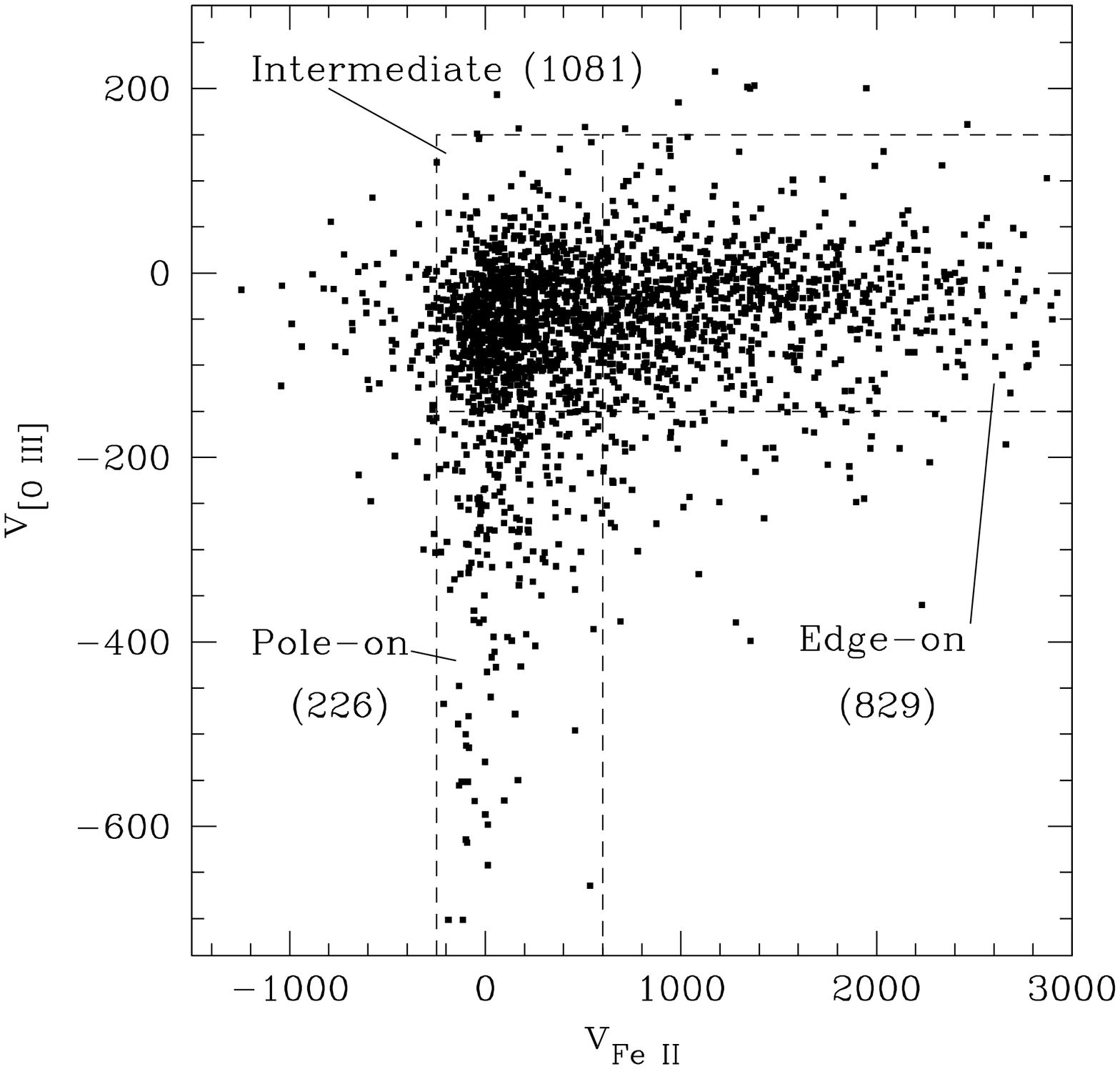}

\caption{The velocity of the [O III] $\lambda$5007 line emission plotted against the velocity of the optical Fe II emission, referenced to the systemic redshift, derived from the [O II] $\lambda$3727 line.  All the velocities are from \citet{hu08}.  The regions marked with dashed lines around the offset velocity tails 
are assumed to be predominantly edge-on and pole-on objects.  The dashed box around the origin
designates a set of intermediate objects.  The number of objects in each region is indicated. }
\label{fig:vo3fe}
\end{figure}

\begin{figure}
\epsscale{1.}
\plotone{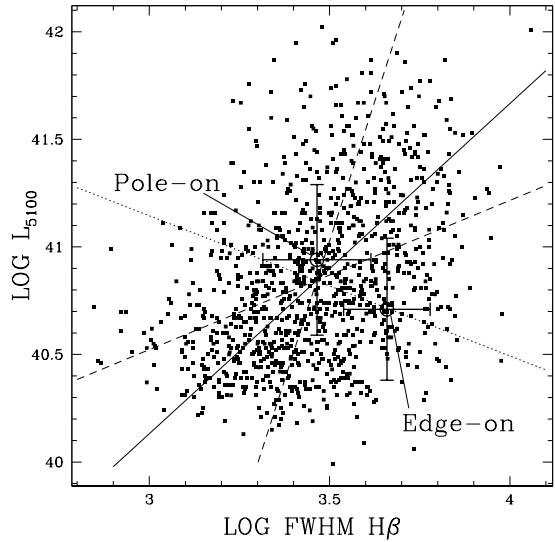}

\caption{Luminosity plotted against H$\beta$ line width for the objects that show no offset velocity
in either Fe II or [O III] and would be intermediate objects if orientation were the only factor.  The two 
large points with error bars show the means and standard deviations
for the subsets of objects identified as edge-on or pole-on because of their large offset velocities.  The 
two dashed lines show two least squares fits to the individual points using each of the two variables
as the independent variable.  The solid line shows the bisector of the two dashed lines. It is clear that the relation between luminosity and line width for the
intermediate objects is inconsistent with the trend indicated by the other two subsets, indicating that 
another factor, Eddington ratio, is important. The dotted line is the adopted dividing line between the two subsets within the part of the sample that shows no offset velocities.}
\label{fig:vo3fe}
\end{figure}

\begin{deluxetable}{lcccc}

\rotate
\tablewidth{0pt}

\tablecaption{Median Properties of Subsets}

\tablenum{1}

\tablehead{\colhead{Property} & \colhead{[O III] outflow} & \colhead{No outflow or inflow} & \colhead{No outflow or inflow} & \colhead{Fe II inflow} \\
\colhead{} & \colhead{High Eddington ratio} & \colhead{High Eddington ratio} & \colhead{Low Eddington ratio} & \colhead{Low Eddington ratio} \\
\colhead{} &  \colhead{Pole-on} & \colhead{Edge-on} & \colhead{Pole-on} & \colhead{Edge-on} }
\startdata
Number of Objects & 226 & 575 & 506 & 829 \\
Log L$_{5100}$ (ergs s$^{-1}$ \AA$^{-1}$) & 40.91 & 40.56 & 41.11 & 40.66 \\
FWHM H$\beta$ (km s$^{-1}$) & 2600 & 2300 & 3800 & 4400 \\
EW [O III] (\AA) & 15.0 & 16.6 & 19.6 & 24.8 \\
EW H$\beta$ (\AA) & 60 & 55 & 73 & 74  \\
EW Fe II (\AA) & 55 & 47 & 48 & 37 \\
FWHM [O III] (km s$^{-1}$) & 771 & 405 & 477 & 432 \\
FWHM [O II] (km s$^{-1}$) & 391 & 389 & 438 & 443 \\
Log M$_{BH}$ (M$_{\sun}$) & 8.20 & 7.88 & 8.59 & 8.40 \\
L$_{bol}$L$_{Edd}$ (Eddington Ratio) & 0.20 & 0.19 & 0.13 & 0.07 \\
\enddata



\end{deluxetable}
\end{document}